\documentclass[preprint]{elsarticle}

\usepackage{amsmath,epsfig}
\usepackage{subfigure}
\usepackage{amssymb,amsfonts}
\usepackage{CJK}
\usepackage{latexsym}
\usepackage[T1]{fontenc} 
\usepackage{amsmath,amssymb,amsthm}
\usepackage{latexsym,graphicx,bbm}
 \usepackage{pdfsync}

\newcommand{\beq}{\begin{eqnarray}}
\newcommand{\eeq}{\end{eqnarray}}

\newcommand{\p}{\partial}

\newcommand{\ba}{\left( \begin{array}}
\newcommand{\ea}{\end{array} \right)}

\begin{document}

\begin{frontmatter}



\title{\boldmath Near integrability of kink lattice with higher order interactions}


\author[a,b]{Yun-Guo Jiang\corref{cor1}}
\ead{jiangyg@sdu.edu.cn}

\cortext[cor1]{Corresponding author}
\author[c]{Jia-Zhen Liu }
\author[d,e]{Song He}
\ead{hesong17@gmail.com}
\address[a]{School of Space Science and Physics,  Shandong University at Weihai,  \\ 264209 Weihai, China}
\address[b]{Shandong Provincial Key Laboratory of Optical Astronomy and Solar-Terrestrial Environment,  Institute of Space Sciences, Shandong University,Weihai, 264209, China}
\address[c]{Department of physics, University of Miami, Coral Gables, FL, 33126, USA}
\address[d]{Max Planck Institute for Gravitational Physics (Albert Einstein Institute)
Am M\"{u}hlenberg 1, 14476 Golm, Germany}
\address[e]{State Key Laboratory of Theoretical Physics, Institute of Theoretical Physics, \\ Chinese Academy of Science, Beijing 100190, P. R. China}

\begin{abstract}
We make use of Manton's analytical method to investigate the force between kinks and anti-kinks at large distances in $1+1$ dimensional field theory. The related potential has infinite order corrections of exponential pattern, and the coefficients for each order are determined. These coefficients can also be obtained by solving the equation of the fluctuations around the vacuum. At the lowest order, the kink lattice represents the Toda lattice. With higher order correction terms, the kink lattice can represent one kind of generic Toda lattice.  With only two sites, the kink lattice is classically integrable. If the number of sites of the lattice is larger than two, the kink lattice is not integrable but is a near integrable system. We make use of Flaschka's variables to study the Lax pair of the kink lattice. These Flaschka's variables have interesting algebraic relations  and non-integrability can be manifested. We also discuss the higher Hamiltonians for the deformed open Toda lattice, which has a similar result to the ordinary deformed Toda.
\end{abstract}


\end{frontmatter}


\section{Introduction}
\label{}




Solitons are non-dissipative objects which occur in many physical models.  In many works, solitonic waves have been found from an integrable system by solving the equations directly. In the other direction, a few studies have constructed many-body systems, and even integrable systems, from soliton objects. Since solitons can be approximated by quasi particles from a dynamical point of view, one can in principle construct integrable models from solitons. This work aims to construct the Toda lattice from the kink lattice.

The interaction between kink and anti-kink is important for scattering processes. Manton proposed an analytical method to calculate such a static force  \citep{Manton:1979,Manton:2004tk,Vachaspati:2006zz}, and found that the force is universal for many scalar field models in 1+1 dimensional field theory. In Manton's work, only the leading order interactions between the kink pair are considered. Recently, we studied the interaction up to second order, and found it to be universal for $\varphi^{4}$, $\varphi^{6}$ and $\varphi^{8}$ models \citep{He:2016}. In this work, we aim to investigate the static force with infinite order corrections. We study the $\varphi^{4}$ theory as an illustration. The coefficients for each order are determined. We find that the dynamics of excitations around the vacuum  determine the interaction pattern. The interaction between the kink pair may play an important role in kink collision phenomena, especially to explain the escape window and the bounce resonance \citep{Moshir:1981,Campbell:1983,Dorey:2011yw,Gani:2014gxa,Gani:2015cda}.

The Toda lattice is one of the most representative and fundamental types of finite dimension Hamiltonian integrable systems \citep{Toda:1967}.
Its integrability was established by Flaschka, Henon and Manakov \citep{ Flaschka:1974,Henon:1974,Manakov:1974}. In our research, the form of the kink lattice Hamiltonian is very similar to the Toda lattice Hamiltonian. Looking only at the leading order interaction, the kink lattice is exactly the same as the Toda lattice. When higher order interactions are included, the kink lattice turns out to be a special deformed Toda lattice. Sawada and Kotera proved that the Toda lattice potential is a unique integrable potential of H\'{e}non type \citep{Sawada:1976}. Higher order terms will therefore break the integrability of the kink lattice.
A kink lattice with higher order corrections become a near or quasi integrable system \citep{Gignoux:2009,Ferreira:2010gh,Ferreira:2016ubj}. We show such near integrability in terms of the Flaschka variables. The ``Lax pair"  has been constructed for the kink lattice and the generalized deformed Toda lattice. This ``Lax pair" can represent the integrable Lax pair of the Toda lattice, and can show how the higher order interactions break the integrability.  We find some non-trivial algebraic relations for the Flaschka variables. Several cases are discussed to show the rich physics of the kink lattice.
The paper is organized as follows. We construct the kink lattice in Section 2.
 Many aspects of the kink lattice are discussed in Section 3. Discussion and conclusions are given in the last section.

\section{The kink lattice \label{KL}}

To construct the kink lattice, we use the $\varphi^4$ kink for illustration. The $1+1$ dimension Lagrangian for $\varphi^4$ theory  reads
\begin{align} \label{BL}
{\cal L}= & \frac{1}{2} \p_{\mu}\varphi \p^{\mu} \varphi - \lambda(\varphi^2-v^2)^2 ,
\end{align}
where $\mu=0,1$ and $\lambda$ is the coupling constant. The Euler-Lagrange equation is written as
\beq \label{EL}
\ddot{\varphi}-\varphi''+ \frac{d V}{d \varphi}=0.
\eeq
The soliton exists in a static condition, and the energy of the system is given by
\beq
E=\int_{-\infty}^{+\infty}(\frac{1}{2} \varphi'^2 \pm V(\phi)) d x \geq \mp \int_{-\infty}^{+\infty} \sqrt{2V(\varphi)} d \varphi,
\eeq
where the equals sign stands for the Bogomol'nyi bound. The BPS equation for the kink is written as
\beq \label{bps}
\varphi'=\pm\sqrt{2V(\varphi)} .
\eeq
The kink solution must interpolate between $v$ and $-v$. Here, we define $(\varphi_{-\infty}, \varphi_{\infty})=(-v, v)$ as the kink, and  $(v, -v)$ as the antikink. The kink solution can be written as
\beq
\varphi(x)= v \tanh \left[ \sqrt{2\lambda} v(x-x_0) \right],
\eeq
where $x_0$ denotes the position of the kink. The antikink solution  can be obtained by replacing $x$ with $-x$.

 The momentum of the system is written as \footnote{The  momentum tensor is defined as $T_{\mu \nu}=\frac{\delta {\cal L}}{\delta \partial \mu \phi} \partial_{\nu}\phi-g_{\mu \nu} {\cal L}$.}
\beq
P= -T_{01}= - \int_{-\infty}^b \dot{\varphi} \varphi' d x.
\eeq
Here we consider the generic case rather than the static case. The classical force is then derived by
\beq \label{eq:force}
F=\frac{dP}{dt}=\left[-\frac{1}{2}(\dot{\varphi}^2 +\varphi'^2 )+ V(\varphi) \right]_{-\infty}^b,
\eeq
which is also valid  for the  motion of the field. We have not considered the static kink solution up to this step.

Now we consider the static interaction between the kink $\varphi_1$ and the antikink $\varphi_2$, whose positions are $x_{01}$ and $x_{02}$ respectively.
$\varphi_1$ and $\varphi_2$ are written as
\begin{align} \label{kak}
\varphi_1= &v\tanh[\sqrt{2 \lambda} v(x-x_{01})], \nonumber \\
\varphi_2=&v\tanh[-\sqrt{2 \lambda} v(x-x_{02})].
\end{align}
We set $R= x_{02}-x_{01}> 0$ to be large but not infinite.
The kink and antikink pair configuration is represented by
\beq  \label{varphi}
\varphi=\varphi_1+\varphi_2-v,
\eeq
 where $v$ is the vacuum at the center of the pair. Now $\varphi$ is independent of time. One can omit the first term in Eq.~(\ref{eq:force}).
At $-\infty$, both $\varphi_1$ and $\varphi_2$ approach the vacuum, and their derivatives are zero. So, the force is related only to the pressure at $b$.
We also set $b$ to be the center of the pair, i.e., $b = \frac{1}{2}(x_{01}+x_{02})$.

 At  point $b$, both $\varphi_1$ and $\varphi_2$ approach the vacuum $v$. One can set
\beq \label{phi12}
\varphi_1 \equiv v + \chi_1, \qquad \varphi_2  \equiv v +\chi_2,
\eeq
where $\chi_{1,2}$ denotes the perturbation field around the vacuum. The potential $V(\varphi)$ can be expanded around the vacuum $v$ as a power series of the perturbation $\chi$\citep{Manton:1979}, i.e.,
\beq \label{V}
V(v +\chi)= \sum_{n=0}^{\infty} \frac{1}{n!} V_{(n)} \chi^{n},
\eeq
where $V_{(n)}=\frac{d^n V }{d^n \varphi}|_{\varphi=v}$. Equation~(\ref{EL}) indicates that $\chi$ satisfies the following equation:
\beq \label{CE}
\frac{d^2 \chi}{ dx^2} -\tilde{m}^2 \chi =\sum_{n=2}^{\infty} \frac{1}{n!} V_{(n+1)} \chi^n,
\eeq
where $\tilde{m}= 2\sqrt{2\lambda} v$ is the mass of $\varphi$. Near the vacuum, $\chi$ can be expanded as
\beq \label{chitt}
\chi(x)=\sum_{k=1}^{\infty} a_k \exp (-k\tilde{m}|x-x_0|).
\eeq
$a_1$ take arbitrary values, and other coefficients can be determined subsequently as
\beq \label{a1a2a3}
a_1, \, a_2=a_1^2 \frac{V_{(3)}}{6 \tilde{m}^2}, \, a_3=a_1^3\left(\frac{V_{(3)}^2}{48\tilde{m}^4}+ \frac{V_{(4)}}{48\tilde{m}^2} \right), \ldots
\eeq
For $a_1=-2v$, it can be verified from Eq~(\ref{a1a2a3}) that $a_2=2v$, $a_3=-2v$, etc. The kink solution enables us to expand $\varphi$ field by Taylor expansion.  One can obtain the expression for $\chi_{1,2}$ from Eq.~(\ref{kak}), i.e.,
\begin{align}  \label{chi12}
\chi_1=&2v \sum_{k=1}^{\infty} (-1)^k e^{-k\tilde{m}(x-x_{01})}, \nonumber \\
 \chi_2=&2v \sum_{k'=1}^{\infty} (-1)^{k'}e^{k'\tilde{m}(x-x_{02})}.
\end{align}
Here we expand the field near the vacuum $v$ at point $b$, where  $\chi_1$ is equal to $\chi_2=2v \sum_{k=1}^{\infty} (-1)^k e^{-k\tilde{m}R/2}$.  Comparing Eq.~(\ref{chitt}) and Eq.~(\ref{chi12}), $\chi$ is equal to either $\chi_1$ or $\chi_2$.  This proves that the coefficients of the Taylor expansion are indeed the solutions of Eq.~(\ref{CE}). The agreement between these two theories is not a coincidence, since the static field equation (\ref{CE}) is just the spatial derivation of the BPS equation (\ref{bps}) for $1+1$ dimensional scalar theory.
Since the static field equation is a second differential, $a_1$ are arbitrary. However, if we consider the first differential BPS equation, $a_1$ can be determined uniquely at the boundary.
The $\chi$ field denotes the perturbation around the vacuum in the theory. The argument here indicates that the static kink configuration can be used to study the dynamics of excitations classically.

Now we consider the static force between the kink and antikink in Eq.~(\ref{eq:force}). The first time-differential term can be ignored for the static condition. Then, the force can be calculated by substituting Eq.~(\ref{varphi}) and Eq.~(\ref{phi12}) into Eq.~(\ref{eq:force}), i.e.,
\beq \label{FT}
F=-\frac{1}{2}(\chi_1'+\chi_2')^2+\sum_{n=0}^{\infty} \frac{V_{(n)}}{n!}(\chi_1+\chi_2)^n.
\eeq
For the $\varphi^4$ potential, $V$ can only be expanded up to the fourth derivation $V_{(4)}$.
Substituting the expression for $\chi_{1,2}$, one obtains the results
\beq \label{FS}
F=8\tilde{m}^2 v^2 \sum_{n=1}^{\infty} \alpha_n e^{-\frac{n+1}{2}\tilde{m}R},
\eeq
where $\alpha_n=-\frac{(-1)^n}{3}(n+2n^3)$. In Eq.~(\ref{FS}), the first term disappears since $\chi_1'+\chi_2'=0$ at point $x=b$. Thus, the force originates completely from the potential term. The sign of each order correction is against the next order, which indicates an alternative attractive or repulsive force. The force in Eq.~(\ref{FS}) includes all orders of corrections. It has been stated that the force should be universal for all species of kink and antikink interaction in one dimension~\citep{Manton:1979}, which needs to be verified for the $\varphi^6$, $\varphi^8$ and sine-Gordon theories.
It is well-known  that there are no interactions between BPS solitons. The analytical method to obtain the static force between kink and anti-kink here considers the point $b$, at which the kink and anti-kink are actually non-BPS. The perturbation around the vacuum represents the non-BPS excitation mode
in the background of kink and anti-kink. These modes actually affect the soliton scattering. Therefore, the analytical method here can be generalized to study other kinds of soliton interactions.


With the normalization $\lambda=1$ and $v=\frac{1}{2\sqrt{2}}$, one can write out the interaction energy according to $F=dU/dR$. The potential $U$ is given by
\beq
U=\sum_{n=1}^{\infty}\beta_n e^{-\frac{n+1}{2}R},
\eeq
where $\beta_n=(-1)^n \frac{2(n+2n^3)}{3(n+1)}$.
One can construct the kink lattice along one line with kink and antikink alternating.  The $i-$th kink or antikink experiences the following force
\beq
\frac{dp_i}{dt}=\sum_{n=1}^{\infty} \alpha_n\left[-e^{-\frac{n+1}{2}(q_{i-1}-q_{i})}+e^{-\frac{n+1}{2}(q_{i}-q_{i+1})}\right].
\eeq
Thus, the kink lattice becomes a kind of  deformed Toda lattice\footnote{Here, the deformed Toda lattice has the same Lagrangian as the kink lattice. The generic Toda lattice in this work refers to  the deformed Toda lattice with generalized coupling coefficients. }. In this way, we have constructed the integrable system from the soliton section directly. The  deformed Toda lattice system produced has not previously been studied, to our knowledge.

The Hamiltonian for the kink lattice can be given by the kink momentum and potential directly. For the nonperiodic (open)  kink lattice, the Hamiltonian is written as
\beq \label{HNP}
H=\sum_{i=1}^{N}\frac{1}{2}p_i^2 +\sum_{i=1}^{N-1}\sum_{n=1}^{\infty}\left( \beta_n e^{-\frac{1+n}{2}(q_i-q_{i+1})}\right),
\eeq
where $N$ is the number of lattice sites. For the periodic (closed) case, the Hamiltonian is given by
\beq  \label{HP}
H=\sum_{i=1}^{N}\frac{1}{2}p_i^2 +\sum_{i=1}^{N}\sum_{n=1}^{\infty}\left( \beta_n e^{-\frac{1+n}{2}(q_i-q_{i+1})}\right), \, q_{N+1}=q_1.
\eeq
In the kink lattice, the periodic condition means to glue the right arm of the $N$th kink to the left arm of the first kink. The vacua must be correctly connected by the kink and antikink. Therefore, $N$ should be an even integer for the periodic kink lattice. However, there is no such constraint for the periodic deformed Toda lattice, $N$ can be any integer.  We will consider the integrability of the kink lattice in the following.

\section{Near integrability \label{AKL}}

Although we have obtained the Hamiltonian of the kink lattice, its integrability needs to be verified.
The Lax pair representation of the system is essential to prove the integrability classically. Flaschka's transformation enables us to give the Lax pair \citep{Flaschka:1974}. Instead of variables $p_i$ and $q_i$, we set new variables to describe the system. For the open Toda lattice, one can set
\begin{align}
a_i\equiv &\sqrt{\sum_{n=1}^{\infty}\left( \beta_n e^{-\frac{1+n}{2}(q_i-q_{i+1})}\right)}, \, i=1, \ldots, N-1 \\
b_i\equiv &p_i.
\end{align}
 In terms of $a_i$ and $b_i$, the Hamiltonian of the open Toda is written as
\beq
H=\frac{1}{2}\sum_{i=1}^{N}b_i^2+ \sum_{i=1}^{N-1}a_i^2.
\eeq
The Lax pair satisfies $\dot{L}=[M,L]$, which are assumed to have the forms
\begin{align} \label{LGT}
L=&\left(\begin{array}{ccccc}
b_1 & a_1 & 0 & \ldots & 0 \\
a_1 & b_2 & a_2 &  & \vdots \\
0 & a_2 & b_3 & \ddots & \vdots \\
\vdots &  & \ddots & \ddots & a_{N-1} \\
0 & \ldots & \ldots & a_{N-1} & b_N \\
\end{array}
\right), \nonumber \\
M=&\left(\begin{array}{ccccc}
0 & -c_1 & 0 & \ldots & 0 \\
c_1 & 0 & -c_2 &  & \vdots \\
0 & c_2 & 0 & \ddots & \vdots \\
\vdots &  & \ddots & 0 & -c_{N-1} \\
0 & \ldots & \ldots & c_{N-1} & 0 \\
\end{array}
\right).
\end{align}
Here $c_i$ are unknown parameters. The evolution of $L$ leads to the equations for $c_i$, i.e.,
\begin{align} \label{solution}
\dot{a_i}=&c_i(b_i-b_{i+1}), \qquad i=1,\cdots,N-1 \\
\dot{b}_1=&-2c_1a_1, \\
\dot{b_i}=&2(c_{i-1}a_{i-1}-c_ia_i), \qquad i=2, \cdots, N-1 \\
\dot{b}_N=&2a_{N-1}c_{N-1}. \label{LS}
\end{align}
The kink lattice system goes back to the Toda system if we only keep the leading $n=1$ term of the potential. In the Toda theory, the solution for $c_i$ is simply written as
\beq
c_i= \frac{\p a_i}{\p q_i}.
\eeq
The Hamiltonian equations of motion of $a_i$ and $b_i$ read
\beq
\dot{a}_i=c_i(b_i-b_{i+1}), \qquad \dot{b}_i= 2c_i(a_{i-1}-a_i).
\eeq
The Hamiltonian equations of motion for $q_i$ and $p_i$ agree with  Eqs.~(\ref{solution}) to (\ref{LS}).
The Poisson bracketss of $a_i$ and $b_i$ are given by
\beq
\{a_i, b_i\}=c_i, \qquad \{a_i, b_{i+1}\}=-c_i, \qquad  i\leq N-1
\eeq
We find that the $(i,i+2)$ component of $[M,L]$ is
\beq a_i c_{i+1}-a_{i+1}c_{i}, \label{constraint}
\eeq which is not zero in general. Thus, the validity of the Lax pair needs the constraint that $a_i c_{i+1}=a_{i+1}c_{i}$.
If one finds a solution for $c_i$ which satisfies the equations of motion in Eq.~(\ref{solution}) and the constraint $a_i c_{i+1}=a_{i+1}c_{i}$, we can claim that the system is integrable. Otherwise, the system is not integrable.

 Flaschka's form indicates that the kink lattice has a generic algebraic structure in the dynamics.
In the kink lattice, $\beta_n$ is determined by $\alpha_n$. The coefficients can be generalized without hindering the equations of motion (in terms of $a_i,b_j$ and $c_k$).  The integer and half integer index in the exponential term can be generalized to an arbitrary real number. However, the physical system should have finite energy, which puts strong constraints on the coefficients. The Hamiltonian of the proposed  most generic open Toda lattice is given by
\beq \label{IHT}
H=\frac{1}{2}\sum_{i=1}^{N}p_i^2+\sum_{i=1}^{N-1}\sum_{n=1}^{\tilde{N}} \beta_n e^{-k_n(q_i-q_{i+1})}.
\eeq
where $k_n$ is positive. $\tilde{N}$ can be infinite when the summation of all terms is convergent. Then, the algorithm of the Flashcka transformation repeats the Lax pair representation as above, setting
\beq
a_i \equiv \sqrt{\sum_{n=1}^{\tilde{N}} \beta_n e^{-k_n(q_i-q_{i+1})}}, \qquad b_i\equiv p_i.
\eeq
The Lax representation does not change for this new deformed Toda lattice.
Flaschka's variables enable us to discuss the integrability of several special cases.

\subsection{Case studies}
First, if we keep only the leading order interaction in the kink lattice, the theory goes back to the Toda theory exactly. The constraint in Eq.~(\ref{constraint}) is satisfied automatically, since
\beq
c_i=-\frac{1}{2}a_i.
\eeq
This shows that the  Toda lattice is integrable, which is a well known result.
The integrals of the motion can be obtained by
\beq \label{HK}
H_k= \frac{1}{k}{\rm Tr} L^k , \quad k=1,2, \cdots, N
\eeq
The first invariant $H_1$ gives the conservation of the momentum. The second invariant $H_2=\frac{1}{2}{\rm Tr}L^2$ is the Hamiltonian. The invariants also satisfy the relation $\{H_i, H_j\}=0$.

Secondly, if we keep only the leading interaction in the generic Toda lattice in Eq.~(\ref{IHT}), the Hamiltonian in Eq.~(\ref{IHT}) becomes
\beq
H=\frac{1}{2}\sum_{i=1}^{N}p_i^2+\sum_{i=1}^{N-1} \beta_{i} e^{-k_i(q_i-q_{i+1})}.
\eeq
which is the Hamiltonian of the inhomogeneous Toda theory. One can check that
\beq \label{IHD}
a_i c_{i+1} -c_i a_{i+1}=\frac{1}{2}a_ia_{i+1}\left( k_i-k_{i+1}\right) \neq  0,
\eeq
which is not zero. Thus, the inhomogeneous Toda lattice is not integrable. Physically, the different couplings between different sites break the integrability. We give the $N=3$ case for illustration.
According to the definition in Eq.~(\ref{HK}), the three variables $H_1,H_2$ and $H_3$ are given by
\begin{align}
H_1=&b_1+b_2+b_3, \\
H_2=&\frac{1}{2} (b_1^2+b_2^2 +b_3^2)+a_1^2 +a_2^2, \\
H_3=&\frac{1}{3}(b_1^3+b_2^3 +b_3^3)+(b_1+b_2)a_1^2+(b_2+b_3)a_2^3.
\end{align}
The Poisson brackets between them are given by
\begin{align}
\{H_1,H_2\}=&0, \\
 \{H_1,H_3\}=&0, \\
  \{H_2,H_3\}=&(k_2-k_1)a_1^2a_2^2.
\end{align}
These relations show that the system has momentum and energy conservation, but we do not have the third integral of motion. $H_3$ is not a conserved integral of motion. The evolution of $H_3$ is corrected by the difference of the couplings. The
 inhomogeneous Toda is not an integral system.


Thirdly, for the case with only two sites, i.e. $N=2$, the deformed Toda theory with high order exponential terms is exact integrable, since there are no constraint conditions any more. For the same reason, the two-site kink lattice is integrable. For a two-site system, we only need two integrals of motion to manifest the integrability. The dynamics of the Lax pair  agrees with the Hamiltonian equations of motion, which are given by
\begin{align} \label{L11S}
\dot{a}_1= c_1 (b_2-b_1),\quad \dot{b}_1=2a_1c_1, \quad \dot{b}_2=-2 a_1c_1.
\end{align}
One can construct two integrals of motion, which are the momentum and the energy of the system.

 A system with $N$ degrees of freedom is super-integrable if it has $2N-1$ independent constants of motion. We can further ask whether the system we are considering is super-integrable \citep{Agrotis:2006}.
Moser proposed to make use of new variables ($\lambda_i, r_i$) to replace the variables ($a_i,b_i$). The function $f(\lambda)$ is defined as:
 \beq  \label{fLambda}
 f(\lambda)=\frac{1}{\lambda -b_2 -\frac{a_1^2}{\lambda -b_1}}.
  \eeq
$f(\lambda)$ can be expanded in a series of powers of $1/\lambda$, i.e.,
\beq
f(\lambda)=\sum_{j=0}^{\infty} \frac{c_j}{\lambda^{j+1}},
\eeq
where $c_j=\sum_{i=1}^{N}r_i^2\lambda_i^j/\sum_{i=1}^N r_i^2$. The corresponding relations between $a_1, b_i$ and $\lambda_1, r_i$ are given by
\begin{align}
a_1^2=&\frac{r_1^2r_2^2(\lambda_2-\lambda_1)^2}{(r_1^2+r_2^2)^2}, \\
b_1=&\frac{r_1^2\lambda_2+r_2^2\lambda_1}{r_1^2+r_2^2}, \\
b_2=&\frac{r_1^2\lambda_1+r_2^2\lambda_2}{r_1^2+r_2^2}.
\end{align}
The equations of motion for  $\lambda_i$  and $r_i$ are
\begin{align}
\dot{\lambda}_i=&0, \\
\dot{r}_i=&-\lambda_i r_i.
\end{align}
This agrees with the Lax pair representation, i.e.,
\begin{align}
L=\left(\begin{array}{cc}
b_1 & a_1 \\
a_1 & b_2
\end{array}\right),
\qquad
M=\left(\begin{array}{cc}
0 & a_1 \\
-a_1 & 0
\end{array}\right).
\end{align}
However, such a Lax pair representation does not agree with the Hamiltonian equations of motion of the deformed Toda, see Eq.~(\ref{L11S}). One has $a_1\neq c_1$. This probably indicates that the deformed Toda is integrable but not super-integrable. In order to prove super-integrability, one needs to represent $c_1$ with $\lambda_i$ and $r_i$, which has not yet been solved.

Fourth, for the $N\geq3$ case, there  is no solution of $c_i$ which satisfies the equations of motion and the constraint simultaneously. For instance, for the $N=3$ case, $c_1$ and $c_2$ are unknown. There are no self-consistent solutions, or, equally, the equation of motion conflicts with the constraint.  The components $(c_i a_{i+1}-a_i c_{i+1})$ are higher order interaction corrections. One can decompose the Hamiltonian in Eq.~(\ref{HNP}) as
\beq
H(q,p)=H_{0}(q,p)+V(q,p),
\eeq
where $H_0$ represents the Hamiltonian for the Toda lattice, and
\[ V(q,p)=\sum_{i=1}^{N-1}\sum_{n=2}^{\infty} \beta_n e^{-\frac{1+n}{2}(q_i-q_{i+1})} \]
denotes the higher order interactions, which can be considered as the perturbation part.
This indicates that the kink lattice is not exactly but ``near" integrable. One can use the canonical perturbation method to study its dynamics \citep{Gignoux:2009}. The ``near" integrable system interpolates between the integrable system and the chaotic system. The kink lattice offers a nice toy model for such a near integrable system. Recently, Ferreira et al. have used the parameter expansion method to study the sine-Gordon kink model for the breathers and wobbles phenomena \citep{ Ferreira:2010gh,Ferreira:2016ubj}. Our exact analytical results for the kink interaction here can also be  used to study such phenomena.
\subsection{Higher Hamiltionians}
For the ordinary Toda lattice, one can construct the higher Hamiltonians by taking the trace of many powers of the Lax matrix, which are all integrable.  The Hamiltonian of the Toda lattice system is given by
\beq
H=\frac{1}{2} {\rm Tr} (L^2).
\eeq
One can construct the systems with higher Hamiltonians as
\beq
H_k=\frac{1}{k} {\rm Tr} (L^{k}),
\eeq
where $k$ is an integer larger than 2.  If $L$ represents the ordinary Toda, then all the systems with higher Hamiltonians are integrable.
For each $k$, one can have a Lax representation, i.e.,
\beq
\dot{L}=[M_k,L],
\eeq
where $M_k$ is a skew-symmetric matrix.

In the following, we will present the higher Hamiltonians for the deformed Toda lattice, which are near integrable systems.
If $L$ represents the deformed Toda lattice in Eq.~(\ref{LGT}), one can also construct the higher Hamiltonian for the generalized Toda lattice. Similar to the generic Toda, systems with higher Hamiltonians are not integrable, but they are near integrable. We find a ``Lax representation" for these near integrable systems. For instance, we consider the $H_3=\frac{1}{3}{\rm Tr}L^3$ system. The $M_3$ matrix can be given by
\beq
M_3= \left(\begin{array}{ccccccc}
0 & \zeta_1   &\eta_1  & 0 & \cdots& \cdots& 0 \\
 -\zeta_1 & 0   & \zeta_2 & \eta_2 & \cdots &\cdots&0\\
 -\eta_1&  -\zeta_2 &0 & \zeta_3 & \ddots &\cdots& 0 \\
  \vdots & \ddots & \ddots & \ddots& \ddots & & \vdots \\
 \vdots &  & \ddots &\ddots& \ddots & \zeta_{n-2} &\eta_{n-2}\\
\vdots &  & & \ddots& -\zeta_{n-2}& 0&\zeta_{n-1}\\
0 & 0&0& \cdots & -\eta_{n-2} & -\zeta_{n-1} & 0\\
\end{array}
\right).
\eeq
 The components are given by
 \begin{align}
 \zeta_i&= c_i(b_i+b_{i+1}),& \qquad k&=1, \ldots, n-1 \\
 \eta_i&= c_ia_{i+1}.&\qquad k&=1, \ldots, n-2
 \end{align}
The Lax representation leads to the  equations of motion, which  are given by
\begin{align}
\dot{a}_i = &-a_{i-1} c_{i-1}a_i + c_i(a_{i+1}^2-b_{i}^2+b_{i+1}^2),\,\nonumber \\
&{\rm for} \qquad k=1, \ldots, n-1 \\
\dot{b}_i =& - 2a_{i-1}c_{i-1} (b_{i-1}+b_i)+2a_ic_i(b_i+b_{i+1}),\nonumber \\
&{\rm for}  \qquad k=1, \ldots, n
 \end{align}
where one also sets $a_0=a_{n}=c_n=b_0=b_{n+1}=0$. Besides that, other non-zero components in the $[M_k,L]$ matrix are proportional to ($a_{i}c_{i+1}-a_{i+1}c_{i}$). We take the $n=4$ case for illustration. One obtains that
\begin{align}
[M_3,L]_{13}&=(b_2+b_3) (a_2 c_1-a_1c_2), \\
[M_3,L]_{14}&=a_3 (a_2 c_1-a_1 c_2), \\
[M_3,L]_{24}&=(b_3 + b_4) (a_3 c_2 - a_2 c_3).
\end{align}
The subscripts $13$, $14$ and $24$ denote the matrix components. One concludes that these new higher Hamiltonian systems are still near integrable systems. For the ordinary Toda case, $a_i \propto c_i$, all these terms disappear. Thus, the Hamiltonian of the kink lattice is near integrable, and the higher Hamiltonians of the kink lattice are still near integrable systems.

\subsection{The closed case}

For representation of the closed Toda lattice,  one can repeat the technique with the generalized Hamiltonian in Eq.~(\ref{HP}). Assume that the system is integrable. The Lax formula can be constructed as \citep{Babelon:2003,He:2016}
\begin{align}
L=&\left(\begin{array}{ccccc}
b_1 & a_1 & 0 & \ldots & \lambda^{-1} a_{N} \\
a_1 & b_2 & a_2 &  & \vdots \\
0 & a_2 & b_3 & \ddots & \vdots \\
\vdots &  & \ddots & \ddots & a_{N-1} \\
\lambda a_{N} & \ldots & \ldots & a_{N-1} & b_N \\
\end{array}
\right),
\end{align}
where $a_N=\sqrt{\sum_{n=1}^{\tilde{N}}\left( \beta_n e^{-k_n(q_N-q_{1})}\right)}$, and $\lambda$ is the spectral parameter. The $M$ matrix is given by \citep{Babelon:2003}
\beq
M=\left(\begin{array}{ccccc}
0 & -c_1 & 0 & \ldots &  \lambda^{-1} c_N\\
c_1 & 0 & -c_2 &  & \vdots \\
0 & c_2 & 0 & \ddots & \vdots \\
\vdots &  & \ddots & 0 & -c_{N-1} \\
-\lambda c_N & \ldots & \ldots & c_{N-1} & 0 \\
\end{array}
\right).
\eeq
The periodic condition is given by $q_{N+i}=q_i$. We list the equations of motion for several cases in the following.

For the $N=2$ case, the system is integrable. The Hamiltonian is given by
\beq
H=\frac{1}{2}(b_1^2+b_2^2) +a_1^2.
\eeq
The equation of motion is given by
\begin{align}
\dot{a}_1=&\frac{c_1}{\lambda}(b_2-b_1)=\lambda c_1(b_2-b_1), \nonumber \\
 \dot{b}_1=&a_1c_1 (\frac{1}{\lambda}+\lambda)=-\dot{b}_2.
\end{align}
$\lambda$ is equal to $\pm1$ for consistency. Thus, the spectral curve reduces to the two-point trivial case. The two integrals of motion denote the total momentum and energy conservation, respectively.

For the $N=3$ case, one obtains the equations of motion
\begin{align}
\dot{b}_i=& 2(a_{i-1}c_{i-1}-a_ic_i), \\
\dot{a}_1=& {c_i} ({b_i} - {b_{i+1}})  -\frac{ 1}{\lambda}({a_{i-1}}  {c_{i+1}}-{a_{i+1}}  {c_{i-1}}) \nonumber \\
          =& {c_i} ({b_i} - {b_{i+1}}) - \lambda({a_{i-1}}  {c_{i+1}}-{a_{i+1}}  {c_{i-1}} ),
\end{align}
where $i=1,2,3$. Here we have considered the periodic conditions. The constraint term $a_ic_{i+1} -c_i a_{i+1}$ appears. For the ordinary periodic Toda, these constraints disappear.  The non-zero constraint  will involve $\lambda$ in the equations of motion. Also,
the consistence requires $\lambda=\pm1$. Thus, the higher order perturbation term  will trivialize the spectral curve. From this evidence, the higher Hamiltonian systems of the deformed Toda lattice are near or quasi integrable systems \citep{Gignoux:2009,Ferreira:2010gh}.

\section{Discussion and Conclusions}
In this paper, we have calculated the effective potential of a kink and anti-kink pair at large separation. The assumption of large distance is essential for Manton's method. The resulting potential contains infinite order corrections of the exponential type. All coefficients for these orders are obtained exactly. Such an effective potential may play an important role in the kink antikink collision test. Many studies have been done to investigate kink scattering by the numerical method \citep{Moshir:1981,Campbell:1983,Dorey:2011yw,Gani:2014gxa,Gani:2015cda}. \citet{Moshir:1981} first studied kink and antikink scattering by solving the relativistic $\varphi^4$ theory. They found a threshold escape velocity of $0.2594$. Below this value, there are also escape windows for the solitons to scatter off after two oscillations. \citet{Campbell:1983} further indicated that the escape windows are related to the orbital frequency of the bound kink-antikink pair and the oscillation field localized in the pair center. The effective theory obtained in this work can be used to study these phenomena, since the perturbation field around the vacuum plays the role of the oscillation field in Campbell et al.'s work. The effective theory from Manton's method may help to explain the escape window more physically, since the higher order corrections in the potential are most related to the frequency and radiation of the soliton scattering.  We observed that the fluctuation of the $\varphi^4$ theory around the vacuum satisfies a second  differential equation, and its solution agrees with the BPS equation. This phenomenon helps us to calculate the quantum theory from the topological section. If we consider the theory in one time and zero space dimensions, the Lagrangian in Eq.~(\ref{BL}) will become an anharmonic oscillator with a double well potential theory \citep{Gildener:1977sm}.  The kink solution here will become the instanton solution. Our effective potential of the kink and antikink can help to study the instanton and anti-instanton contribution to the kernel.  It can be expected that discrete energy levels for the anharmonic oscillator are related to the instanton configurations.
Manton's method here can also be used to study kink dynamics in other theories, for instance the kink solutions (which are confined monopoles on the vortex string) in SQCD theories \citep{Shifman:2004dr,Eto:2011cv,AlonsoIzquierdo:2008rk}. Several articles have studied the dynamics of such kinks \citep{Arai:2014hda,Harland:2009mf,Tong:2003ik}. It was found that the potential for kink interactions has a similar exponential pattern. The procedure here to calculate the high order potentials and construct the kink lattice can also be applied to the massive sigma model. One can expect that these confined monopoles form a near integrable system. All these predictions should be tested in the future.

The kink can be considered as a pseudoparticle, whose dynamics shows rich structure. We have presented the effective total Hamiltonian for the kink lattice, and generalized the kink lattice to the generic deformed Toda lattice. Keeping only the leading order potential, the kink lattice is exactly the Toda lattice, which is integrable. If higher order terms are included, the kink lattice and the deformed Toda lattice are near integrable, except for the two sites case. In terms of Flaschka's variables, we studied the Lax representation of the kink lattice. Although the potential has infinite correction terms, the Lax equation shows a simple algebraic structure. The integrability is broken by the higher order corrections. This shows that the kink lattice is a near integrable system. It has been stated that the Toda potential is the unique potential for integrability \citep{Sawada:1976}, but this theorem indicates that the inclusion of higher order corrections will break the integrability. We found that the two-site kink lattice is integrable, which is mostly related to the breather phenomenon of the kink and antikink. The two-site kink lattice is most probably not super-integrable, i.e., there are no $2N-1$ integrals for the kink lattice, and further study is needed to verify this point. The higher Hamiltonians for the kink lattice were constructed, and they are all near integrable systems. The evolution of near integrable systems is also interesting, since they will evolve from integrable to chaotic systems.
The kink lattice is a nice model for studies of near integrable systems.

\vskip 0.5cm
{\bf Acknowledgement}
\vskip 0.2cm
We are grateful to Toshiaki Fujimori, Muneto Nitta, and Kesukei Ohashi for useful conversations and correspondence.  The work of Y. G. Jiang is supported by Shandong Provincial Natural Science Foundation (No.ZR2014AQ007) and National Natural Science Foundation of China (No. 11403015 and U1531105). S. He is supported by Max-Planck fellowship in Germany and the National Natural Science Foundation of China (No.11305235).



\end{document}